\documentclass[conference]{IEEEtran}
\IEEEoverridecommandlockouts
\usepackage[utf8]{inputenc} % allow utf-8 input
\usepackage[T1]{fontenc}    % use 8-bit T1 fonts
\usepackage{hyperref}       % hyperlinks
\usepackage{url}            % simple URL typesetting
\usepackage{booktabs}       % professional-quality tables
\usepackage{amsfonts}       % blackboard math symbols
\usepackage{nicefrac}       % compact symbols for 1/2, etc.
\usepackage{microtype}      % microtypography
\usepackage{lipsum}		% Can be removed after putting your text content
\usepackage{graphicx}
\usepackage{doi}
\usepackage{balance}

\usepackage{graphicx}
\usepackage{float}
\usepackage{amsmath}
\usepackage{setspace} 
\usepackage{units}
\usepackage{bmpsize}
\usepackage{array}
\usepackage{empheq}
\usepackage{algorithm}
\usepackage{algpseudocode}
\usepackage{enumitem}   
\usepackage[caption=false]{subfig}

\newtheorem{thm}{Theorem}[section]

\newtheorem{rem}[thm]{Remark}

\begin{document}

\title{Sequential Structure and Control Co-Design of Lightweight Precision Stages with Active Control of Flexible Modes}

\author{
\IEEEauthorblockN{Jingjie Wu}
\IEEEauthorblockA{Walker Department of Mechanical Engineering\\
	The University of Texas at Austin\\
	Austin, TX, 78712 \\
	\texttt{wujingjie@utexas.edu} }
\and
\IEEEauthorblockN{Lei Zhou}
\IEEEauthorblockA{Walker Department of Mechanical Engineering\\
	The University of Texas at Austin\\
	Austin, TX, 78712 \\
	\texttt{lzhou@utexas.edu} }
}

\maketitle

\begin{abstract}
Precision motion stages are playing a prominent role in various manufacturing equipment. The drastically increasing demand for higher  throughput in  integrated circuit (IC) manufacturing and inspection calls for the next-generation precision stages that have light weight and high control bandwidth simultaneously. In today's design techniques, the stage's first flexible mode is limiting its achievable control bandwidth, which enforces a trade-off between the stage's acceleration and closed-loop stiffness and thus limits the system's overall performance.  To overcome this challenge, this paper proposes a new hardware design and control framework for lightweight precision motion stages with the stage's low-frequency flexible modes actively controlled. Our method proposes to minimize the resonance frequency of the controlled mode to reduce the  stage's weight, and to maximize that of the uncontrolled mode  to enable high control bandwidth. In addition,  the proposed framework determines the placement of the actuators and sensors to maximize the controllability/observability of the stage's controlled flexible mode while minimizing that of the uncontrolled mode, which effectively  simplifies the controller designs. Two case studies are used to evaluate the effectiveness of the proposed framework. Simulation results show that the stage designed using the proposed method has a weight reduction of more than 55\% compared to a baseline stage design. Improvement in control bandwidth was also achieved.  These results demonstrate the effectiveness of the proposed method in achieving lightweight precision positioning stages with high acceleration, bandwidth, and precision.

\end{abstract}

% keywords can be removed

\section{Introduction}

\begin{figure}[t!]
\centering
\subfloat{
\includegraphics[width =0.95\columnwidth, keepaspectratio=true]{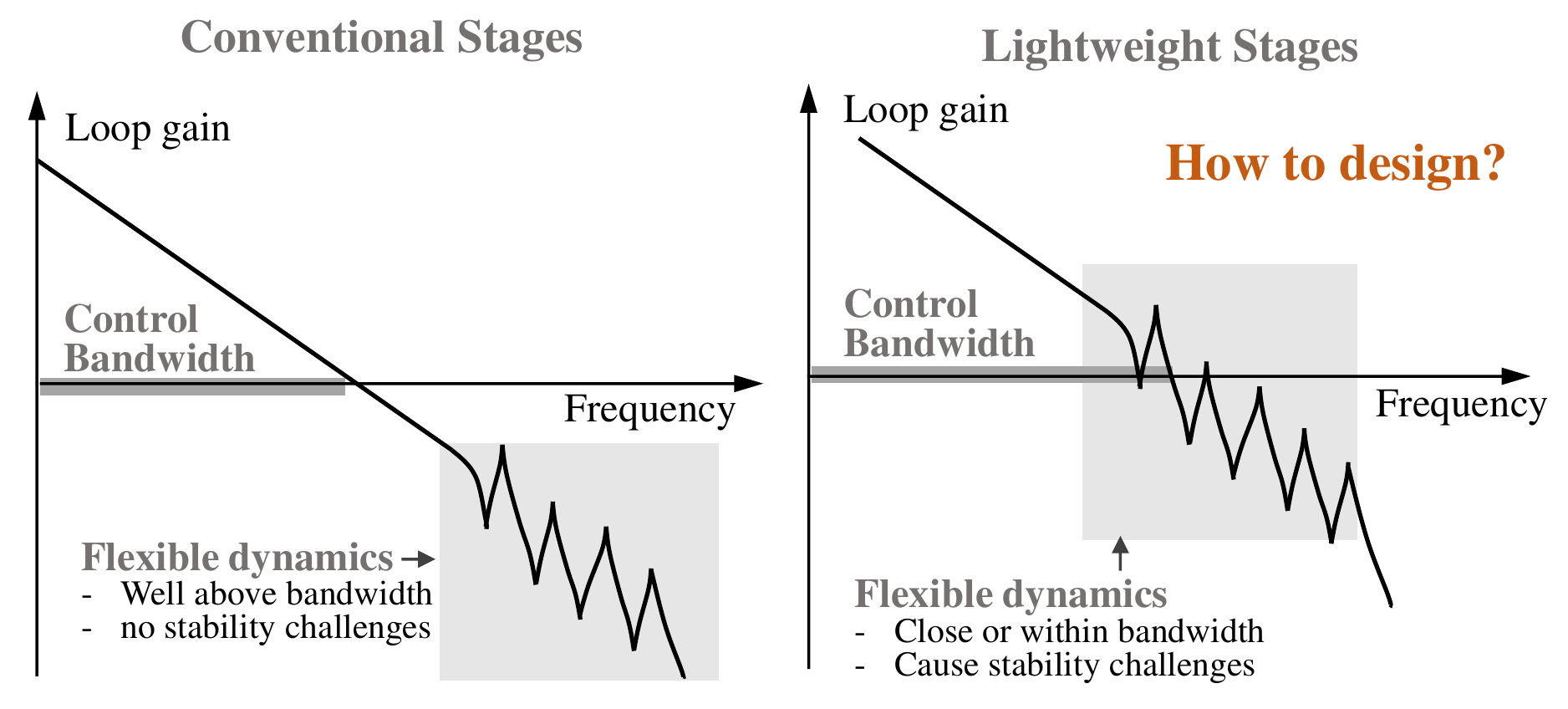}}
\vspace{-2mm}
\caption{Design challenge of lightweight precision positioning stages.}
\vspace{0mm}
\label{fig:research_question}
\end{figure}

High-precision positioning stages are playing a critical role in a wide range of manufacturing and inspection tools such as  photolithography scanners \cite{butler2011position} and MEMS inspection systems \cite{albero2011micromachined}. The drastically growing demand for higher throughput in semiconductor manufacturing necessitates the next-generation precision motion stages with higher acceleration capability while maintaining excellent positioning accuracy and high control bandwidth~\cite{oomen2013connecting}. Creating new lightweight precision positioning stages is critical to achieve this goal. 
%Current stages are designed to have substantially high first resonance frequency above the desired control bandwidth as shown in Fig.~\ref{fig:design_motivation}a to facilitate the controller design, leading to a larger stage weight and limitation on acceleration capability. 
However, as the stage's weight reduces, its structural resonance frequencies will decrease to near or even within the control bandwidth (Fig.~\ref{fig:research_question}), which limits the stage's  control bandwidth and  positioning accuracy, and can even cause stability challenges \cite{oomen2018advanced}. 

\begin{figure}[t!]
\centering
\subfloat{
\includegraphics[trim={0mm 0mm 0mm 0mm},clip,width =0.9\columnwidth, keepaspectratio=true]{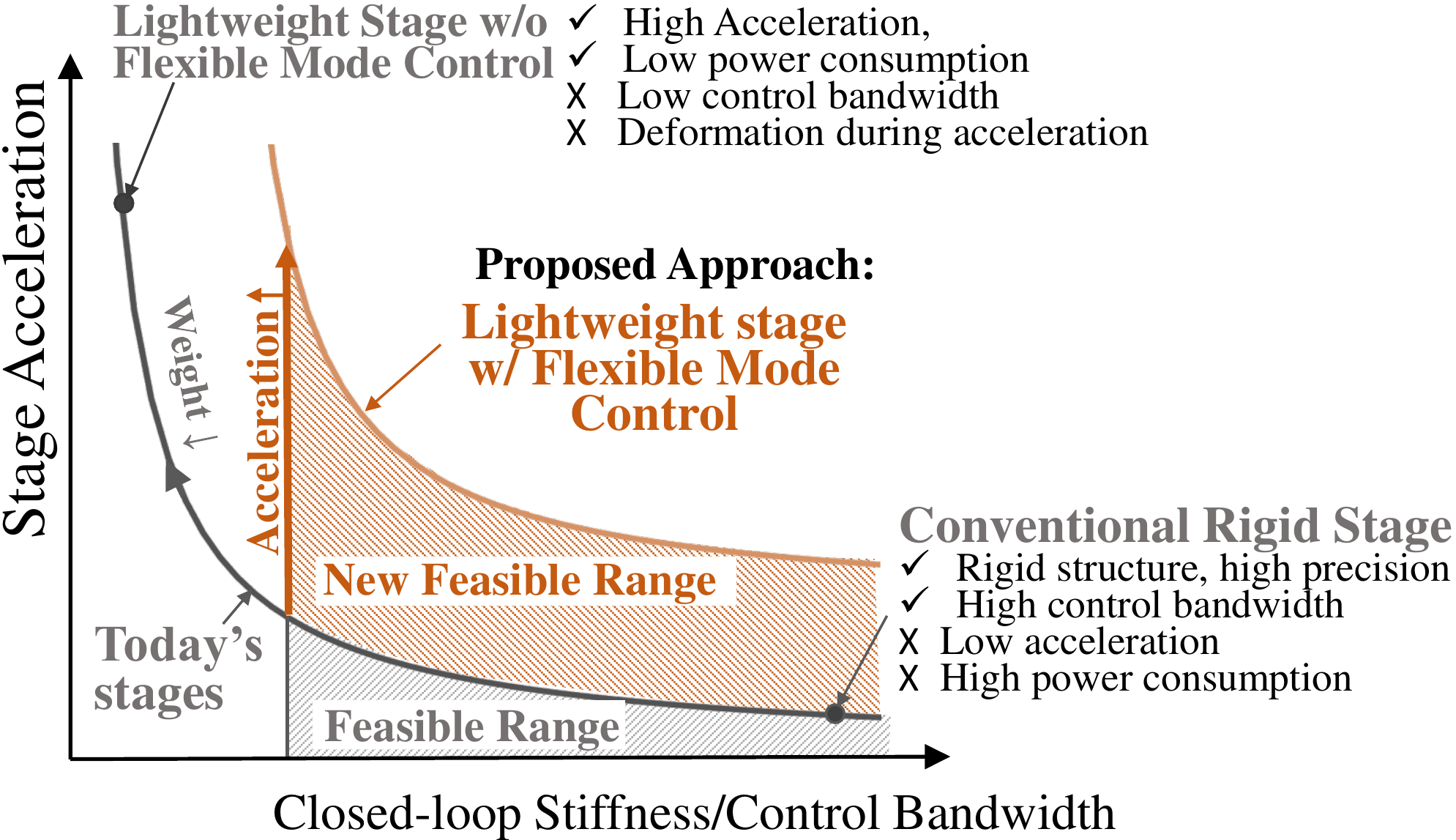}}
\vspace{-2mm}
\caption{Illustration of acceleration and bandwidth trade-off in today's precision positioning systems and motivation for the proposed lightweight stage with flexible mode control.}
\vspace{-6mm}
\label{fig:motivation}
\end{figure}

In the past decade, a number of research and engineering efforts have studied the design and control for lightweight precision positioning stages. For example, Laro~et al.~\cite{laro2010design} presented an over-actuation approach to place actuators/sensors at the stage's nodal locations to prevent the flexible dynamics from being excited by the feedback loops. Oomen et al.~\cite{oomen2013connecting} proposed a system identification and robust control framework for wafer stages, which provides a systematic approach to create controller designs for stages exhibiting low-frequency flexible dynamics. Although effective, these studies mostly investigate the motion control for flexible stages, and the synergy between the structure design and controller design is not fully exploited.  In recent years, the hardware-control co-design, or control co-design (CCD) \cite{garcia2019control}, has been studied for the lightweight precision positioning stages, aiming at enabling a synergistic structure-control design method for precision positioning stages. For example,  Van der Veen et al. \cite{van2017integrating} studied the integrated topology and controller optimization for a simple 2D motion stage structure. Delissen et al.~\cite{delissen2020high} presented a topology-optimized wafer stage fabricated via metal additive manufacturing. In a recent study, Wu et al. \cite{JingjieCoDesign} presented a nested CCD formulation of for lightweight precision stages with controller design constraints explicitly considered. Despite these advances, we make a key observation that in these prior lightweight precision stages designs, the first resonance frequency of the stage structure sets an upper limit for the achievable control bandwidth. This fact enforces a fundamental trade-off between the stage's bandwidth and acceleration as illustrated in Fig.~\ref{fig:motivation}. Fundamental advances in the stage's mechatronic design must be made to break this trade-off and thus enable stages with improved overall performance.

\begin{figure}[t!]
\centering
\subfloat{
\includegraphics[trim={0mm 0mm 0mm 0mm},clip,width =0.6\columnwidth, keepaspectratio=true]{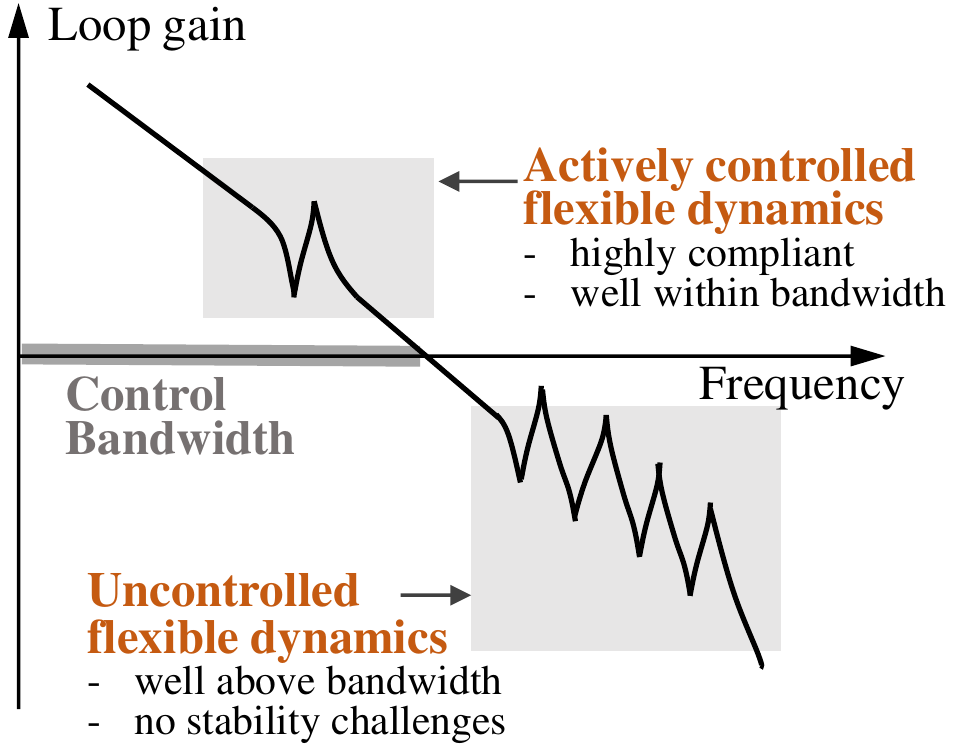}}
\vspace{-2mm}
\caption{Illustration of the  proposed lightweight stage design with active control for flexible modes.}
\vspace{-6mm}
\label{fig:proposed_cartoon}
\end{figure}

Aiming at overcoming the aforementioned trade-off and thus creating new lightweight stages that can simultaneously have high acceleration and high closed-loop stiffness, this paper presents a sequential structure and control design framework where the low-frequency flexible modes of the stage are under active control. This approach has been explored in van Herpen et al.~\cite{van2014exploiting} where additional actuators and sensors are introduced for a lightweight stage to enhance the control bandwidth. However, in~\cite{van2014exploiting}, the control for flexible dynamics is not considered in the stage's structural design phase, which limits the achievable performance. 
In our work, to facilitate the controller design, we propose to minimize the resonance frequency of the stage's mode being controlled and to  maximize the resonance frequency of the uncontrolled mode. The target control bandwidth of the stage is in between the resonance frequencies, as shown in Fig.~\ref{fig:proposed_cartoon}. We envision that this formulation will 
remove material in the stage's structure to allow compliance in the actively-controlled modes thereby breaking the trade-off in lightweight stages, as shown in Fig.~\ref{fig:motivation}. 
With the stage's structure designed, we further propose to use an optimization method to compute the best actuator/sensor placement. Our hypothesis is that maximizing the controllability/observability of the actively-controlled flexible modes while minimizing that of the uncontrolled modes will deliver the best positioning performance with reasonable control signal magnitude. 
%Fig.\ref{fig:design_motivation}c with the following key features: (a) explicitly forcing the first $n$ resonance frequencies within the control bandwidth and actively control them while constraining the rest uncontrolled modes to be above the bandwidth (b) considering hardware parameter and actuator/sensor placement optimization in a unified framework (c) incorporating constraints on closed-loop sensitivity peak to guarantee robustness and (d) low computational cost. The proposed framework is evaluated in two case studies, including a simple stage structure and a practical stage model with extra actuator constraints. 
Two case studies are simulated to evaluate the effectiveness of the proposed approach, where a stage weight reduction of $>55\%$  is demonstrated  compared to a baseline case. These results demonstrate the potential of the proposed lightweight precision stage design framework.

The rest of the paper is organized as follows. Section~\ref{sec:statement} describes the problem statement. Section~\ref{sec:algorithm} presents the proposed design framework for the lightweight precision positioning stage. Section~\ref{sec:simulation} shows the simulation evaluations with two case studies.  Conclusion and future work are summarized in Section~\ref{sec:conclusion}.

%%%%%%%%%  Problem Statement %%%%%%%%%%%%%%%%%%%%%%%%%%%%%%%%%
\section{Problem Statement} \label{sec:statement}
The dynamics of a precision positioning stage considering its flexible structural behaviors can be described by  
\vspace{-2mm}
\begin{align} \label{eqn:mech_EOM_1}
\begin{split}
M(\theta_p)\ddot{x}+D(\theta_p)\dot{x}+K(\theta_p)x &= B(\theta_p,\theta_a)u,
\\
    y &= C(\theta_p,\theta_s)x,
\end{split}
\vspace{-2mm}
\end{align} 
where $x$ is a vector of state variables of both rigid-body displacements and flexible displacements in the modal coordinate, $M$, $D$, $K$ are the mass, damping and stiffness matrices, respectively,  $u$ is the vector of control signals, $y$ is a vector of measurement signals, $B$ is the input matrix which maps the control input $u$ to corresponding states, $C$ is the output matrix which maps state variables to measurements, 
$\theta_p$ is a vector of stage's geometric design parameters, and $\theta_a$, $\theta_s$ are the vectors of actuator and sensor locations, respectively. 

The design optimization problem for a lightweight precision  stage described by \eqref{eqn:mech_EOM_1} aims at finding a set of hardware design parameters $\theta_p$, $\theta_a$, and $\theta_s$ and a controller design that can minimize the stage's weight while maximizing the control bandwidth, meanwhile satisfying certain robustness criteria. 

% first find a set of hardware parameters $\theta$ for the stage structure that minimizes the total weight while satisfying certain modal constraints and then optimize the actuator and sensor locations $b$ and $c$ and synthesis a controller in order to maximize the closed-loop control bandwidth with reasonable robustness.

%%%%%%%%%  Sequential Co-design Framework %%%%%%%%%%%%%%%%%%%%%%%%%%%%%%%%%
\section{Sequential Hardware and Control Optimization Framework} \label{sec:algorithm}

This section presents a sequential framework of designing the hardware and controller for lightweight stages with their low-frequency flexible modes actively controlled. In the first step, an optimization problem that determines the stage's geometric parameter is formulated to facilitate the active control for the stage's low-frequency flexible modes. In the second step, an optimization is performed to determine the location of actuators and sensors.  Finally, feedback controllers are synthesized for the designed stage to control the stage's motion as well as the low-frequency flexible modes. The three steps are introduced in detail in the following sections.

% Our central hypothesises are a) \lei{todo:intro on res frequencies} and b) \lei{todo:method on controllability/observability}

%as a means to break the fundamental trade-off in today's precision positioning stage design techniques, this work proposes 

% This section introduces the CCD problem formulation and a proposed sequential algorithm  for the lightweight motion systems in Section~\ref{sec:statement}. Fig.~\ref{fig:flowchart} shows an overview of

% \begin{figure}[t!]
% \centering
% \subfloat{
% \includegraphics[trim={0mm 0mm 0mm 0mm},clip,width =1\columnwidth, keepaspectratio=true]{Figures/Design_motivation.png}}
% \vspace{-2mm}
% \caption{Design Motivations}
% \vspace{-6mm}
% \label{fig:design_motivation}
% \end{figure} 

\subsection{Stage Geometry Design Optimization}   \label{sec:shape_design}

% \lei{this paragraph looks like background and not our framework. Consider move to introduction section.} 
% In conventional precision positioning stage design techniques, as the stage's weight reduces, the structural resonance frequencies tend to decrease to near the control bandwidth; in this case, the structural dynamics of the stage can be excited due to the stage's acceleration, which obstructs the intended accuracy and can even cause stability challenges \cite{}. Consequently, in prior precision stage designs, the flexible modes of the stage are typically designed to be substantially higher than the stage's control bandwidth to maintain positioning accuracy and control robustness. There are many research investigating the approach to exploit and decrease the gap between control bandwidth and $1st$ vibration natural frequency, e.g., via CCD framework \cite{JingjieCoDesign} \jingjie{cite the previous ACC paper but Idk where to find} \lei{for paper like this you need to write your own bibitem}. However, the weight of the stage is still limited by the first resonance frequency to be high.

% The main challenge in the design and control for lightweight precision positioning systems is that the stage's stiffness can reduce as the weight reduces. Consequently, the stage can undergo structural deformations as it accelerates, which obstructs the intended positioning accuracy. To overcome the aforesaid challenge, in this work, we 

In a lightweight precision stage with active control for low-frequency flexible modes, the stage's geometry design optimization is formulated as 
\begin{align}  \label{eqn: shape_opt}
\begin{split}
    \min_{\theta_p}~~~&{J_m}(\theta_p),
\\
    \mathrm{s.t.}~~~&\omega_i  \leq \omega_{low}, ~~~~~~~ i=1,...,n
\\
    & \omega_j \geq \omega_{high}, ~~~~~ j=n+1,...,m 
\\
    & \theta_{p, min} \leq \theta_p \leq \theta_{p, max}. 
\end{split}    
\end{align}
Here, the objective function $J_m$ represents the stage's weight, $\theta_p$ is a vector for the stage's geometric parameters,  $\omega_i$ is the $i$-th modal frequency with its corresponding vibration mode  actively controlled, and $\omega_j$ is the $j$-th resonance frequency where the corresponding mode shape is not controlled. $\omega_{low}$ is the upper bound for the actively-controlled resonance frequencies, and $\omega_{high}$ is the lower bound for the uncontrolled resonance frequencies.  $\theta_{p, min}$ and $\theta_{p, max}$ are the lower and upper bounds for the stage's geometric parameter, respectively. 

With the stage structure design optimization formulation~\eqref{eqn: shape_opt}, the stage's flexible modes under active control are having resonance frequencies below  $\omega_{low}$, and that of the uncontrolled modes are beyond $\omega_{high}$. Such an optimization process can enforce material removal in the stage's structure to allow for compliance in the actively-controlled flexible modes, and add material to stiffen the uncontrolled modes.%, thereby enhancing the stage's structural stiffness in closed-loop. 

%Furthermore, this shape design concept can be extended as we bound the first $n$ vibration frequencies by $\omega_{low}$ from above and the rest of $r$ higher frequencies starting from $n+1$ mode by $\omega_{high}$ from below. Thus, the weight of the stage can be further reduced while we have to actively control the first $n$ vibration modes. 
% \jingjie{the para below may be moved to the beginning of the case study simulation}
% However, the more low-frequency modes we want to constrain in the design and then actively control, the more complicated stage topology, more actuator/sensor pairs and more computationally expensive actuator/sensor placement procedure have to be used. As a result, we only consider the case where $n = 1$ in this paper to show the feasibility and effectiveness of our framework.
\vspace{-2mm}
\begin{rem}\label{rem:1}
The selection of $\omega_{low}$ and $\omega_{high}$ are highly important and determine the system's dynamic behavior.
The system's target control bandwidth must be between $\omega_{low}$ and $\omega_{high}$, and $\omega_{high}$ sets the new upper bound for the achievable control bandwidth for the lightweight precision stage with actively controlled flexible modes, as illustrated in Fig.~\ref{fig:motivation}. 

% Ideally, we would like to have $\omega_{low}$ as low as possible while having $\omega_{high}$ as high as possible for best closed-loop performance; however such design selections may not be feasible for the stage structure. 
To facilitate controller design while maintaining design feasibility, the values of $\omega_{low}$ and $\omega_{high}$ need to be selected according to the target control bandwidth, for example $\omega_{low}\sim\frac{1}{2}\times \omega_{bw}$ and $\omega_{high}\sim5\times\omega_{bw}$, where $\omega_{bw}$ is the target bandwidth. This method, although robust, may lead to a relatively conservative stage design. To fully evaluate the feasible design range in Fig.~\ref{fig:motivation}, the value of $\omega_{high}$ needs to be swept while considering the actuator/sensor positioning, which will be introduced in Section~\ref{sec:case_study_2}. 
\end{rem}

\subsection{Actuator and Sensor Placement}\label{sec:act_sen_placement}

The actuator and sensor placement optimization problem for the proposed lightweight stage with active flexible mode controlled can be formulated as
\begin{align}  \label{eqn:act_opt}
\max_{\theta_a\in{D_a}}J_a(\theta_a)  = \sum_{i=1,...,n}W_{pi}(\theta_a) - \gamma\sum_{i=n+1,...,m} W_{pi}(\theta_a),
\end{align}
\vspace{-4mm}
\begin{align}  \label{eqn:sen_opt}
\max_{\theta_s\in{D_s}}J_o(\theta_s)  = \sum_{i=1,...,n}W_{oi}(\theta_s) - \gamma\sum_{i=n+1,...,m}W_{oi}(\theta_s),
\end{align}
where $\theta_a$ and $\theta_s$ are vectors of actuator and sensor placement parameters, respectively; $D_a$ and $D_s$ are the design domains for actuator/sensor locations,  and $\gamma$ is a positive user-defined weighting constant. $W_{pi}$ and $W_{oi}$ are the controllability and observability grammians of $i$-th flexible mode, respectively, which can be calculated as 
% \begin{small}
\begin{align}
    W_{pi} = \frac{\| \phi_i(\theta_a)^\top B_a(\theta_a)\|_2^2}{4\zeta_i\omega_i},\; %\label{eq:W_p} \;
    W_{oi} = \frac{\| C_s(\theta_s)^\top \phi_i(\theta_s)\|_2^2}{4\zeta_i\omega_i}\label{eq:W_o},
\end{align}
% \end{small}
\noindent where $\phi_i$ is the mass-normalized mode shape of $i$-th flexible mode, $B_a$ and $C_s$ are the force and measurement assembling matrices, $\zeta_i$ is the modal damping ratio, and $\omega_i$ is the $i$-th resonance natural frequency. The controllability/observability grammians $W_{pi}$ and $W_{oi}$ quantitatively evaluate the controllability/observability of the corresponding flexible mode in the control system, 
which will reflect on the peak resonance magnitude in the system's frequency response.

With actuator/sensor placement optimization formulation in \eqref{eqn:act_opt} and \eqref{eqn:sen_opt}, our goal is to maximize the controllability/observability for the actively-controlled modes to reduce the required controller gain, and to minimize those of the uncontrolled modes to reduce their coupling with the control systems. The value of $\gamma$ provides a trade-off between the two design goals: a low value in $\gamma$ emphasizes reducing the needed controller gain for  actively-controlled modes, and a high value in $\gamma$ emphasizes reducing the cross-talk between uncontrolled modes and controlled modes.

\subsection{Feedback Control Design}

\begin{figure}[t!]
\centering
\subfloat{
\includegraphics[trim={0mm 0mm 0mm 0mm},clip,width =1\columnwidth, keepaspectratio=true]{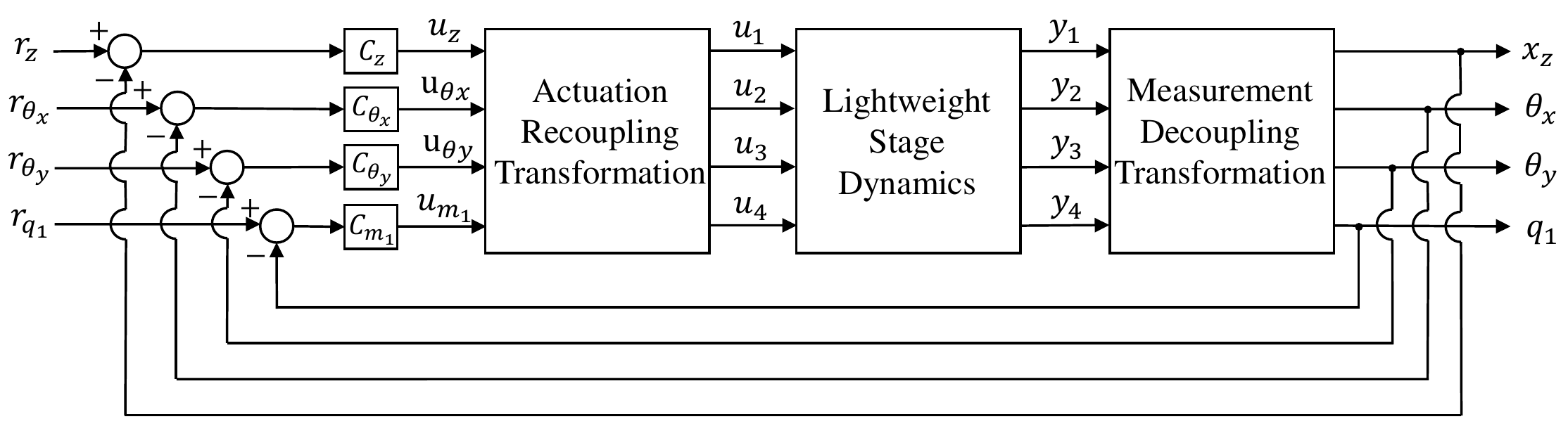}}
\vspace{-2mm}
\caption{Control block diagram for the lightweight precision positioning stage with model decoupling. }
\vspace{-6mm}
\label{fig:control_diagram}
\end{figure}

\begin{table}[t] 
\centering
\vspace{4mm}
\caption{Controller parameters  \cite{butler2011position}. }
\vspace{-4mm}
\label{table:PID_para}
    \begin{center} \begin{small}
        \begin{tabular}{ p{1.2cm} p{4.25cm}  p{1cm} }
        \hline
        Parameter & Description & Typical Value \\
        \hline
        $\omega_{bw}$  & Desired bandwidth [rad/s]  & --  \\
        
        % $m$    &  Stage mass [\rm{kg}]   & --  \\
        
        % $J$ & Stage moment of inertia [$\rm{kg\cdot m^2}$]  &  -- \\
        
        $\alpha$ & PID frequency ratio  &   0.3 \\
        
        $K_p$  &    Proportional gain  & --   \\
        
        $\omega_i$  & Integrator frequency  & $  \omega_{bw}/ \alpha^2$  \\
        
        $\omega_d$  & Differentiator frequency  & $\omega_{bw}/\alpha$ \\
        
        $\omega_{lp}$  & Lowpass filter frequency &  $  \alpha \omega_{bw}$  \\
        
        $z_{lp}$  & Lowpass filter damping ratio  & 0.7\\
        \hline

        \end{tabular}
    \end{small} \end{center}
\vspace{-8mm}
\end{table}

\begin{figure*}[t!]
\centering
\subfloat{
\includegraphics[trim={0mm 0mm 0mm 0mm},clip,width =2\columnwidth, keepaspectratio=true]{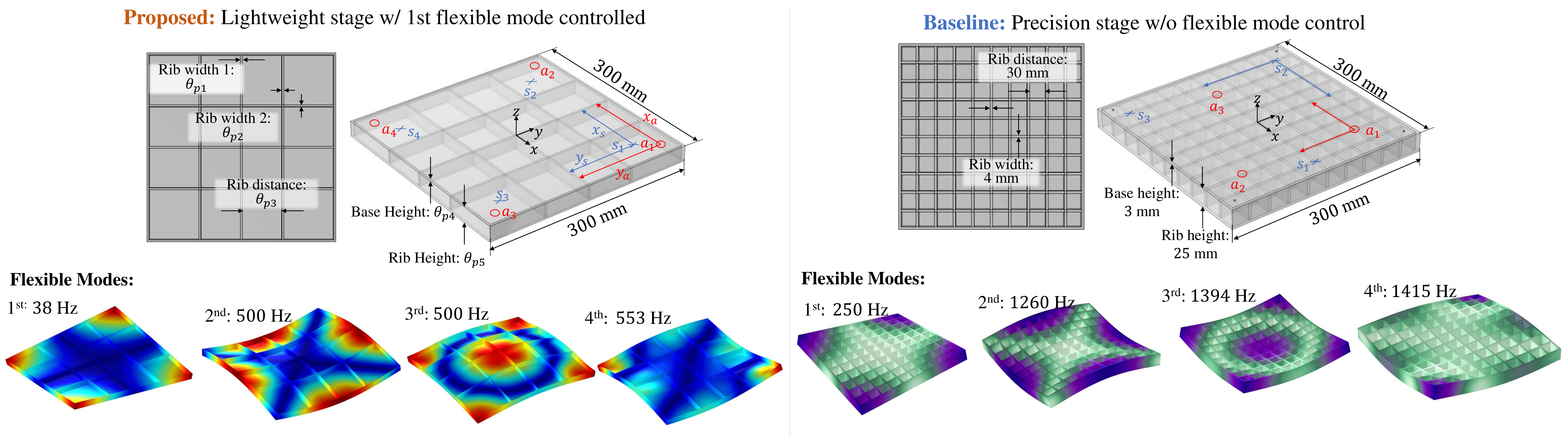}}
\vspace{-2mm}
\caption{Case study \#1: proposed and baseline stage parameter definition and resultant dynamics.}
\vspace{-6mm}
\label{fig:case_1}
\end{figure*}

With the stage's structure and actuator/sensor locations determined, the plant dynamics of the stage can be found. Feedback controllers can be designed for each degree of freedom (DOF) to enable precision positioning and disturbance rejection. 
Figure~\ref{fig:control_diagram} shows a block diagram for the control loop for a lightweight stage with three rigid-body DOFs and one flexible mode under active control. 
Here, the lightweight stage plant dynamics $P:u\to y$
can be obtained from solving  \eqref{eqn: shape_opt}, \eqref{eqn:act_opt}, and \eqref{eqn:sen_opt}. The sensor measurements $y$ are transformed to individual DOFs via a measurement decoupling transformation. Four single-input, single-output (SISO) feedback controllers can then be designed for four decoupled channels %: $u_z\to x_z$ , $u_{\theta x}\to \theta_x$, $u_{\theta y}\to \theta_y$ and $u_{m_1}\to q_1$, 
assuming the cross-coupling between different DOFs is negligible. For each DOF, a fixed-structure SISO controller is selected following reference \cite{franklin2002feedback} as
\begin{align}
\begin{split}   \label{eqn:PID}
    C_k(s) = K_p\Big(\frac{s+\omega_i}{s}\Big)\Big(\frac{s}{\omega_d}+1\Big)\Big(\frac{\omega_{lp}^2}{s^2+2z_{lp}\omega_{lp}s+\omega_{lp}^2}\Big),
\end{split}
\end{align}
where the controller parameters are described in Table~\ref{table:PID_para}. This controller design follows reference \cite{butler2011position, ding2020optimal} where all the controller parameters except the controller gain can be determined by a target control bandwidth $\omega_{bw}$. This approach effectively simplifies the parameter tuning process. %In addition, by changing the parameter $\alpha$, we can tune the trade-off between closed-loop robustness and performance. In this work, a typical value  of $\alpha = 0.3$ is selected following prior works in motion control \cite{ding2020optimal}. 
The proportional gain $K_p$ and the target bandwidth are determined such that the control bandwidth is maximized while satisfying a robustness criteria\cite{ortega2004systematic} of
\begin{align}  \label{eqn:robustness}
 \| S_k(s)\|_{\infty} \leq 2, k = 1, ..., n, 
\end{align}
where $S_k(s)$ is the closed-loop sensitivity function of the $k$-th channel as $S_k = (I-G_k C_k)^{-1}$. 
% The controller design can be done by tuning the parameter $\omega_{bw}$ for each channel. 
With the control effort signals $u_k$ for each channel computed, an actuation recoupling transformation is used to map the control signals to individual actuators.

%%%%%%%%%  Simulation results %%%%%%%%%%%%%%%%%%%%%%%%%%%%%%%%%
\section{Simulation Evaluation}  \label{sec:simulation}

% To evaluate the potential of the proposed method for enhancing acceleration while maintaining high control performance and closed-loop structural stiffness in precision positioning stages, this section presents the numerical evaluation of two case studies designed using the proposed sequential CCD approach. 
Two case studies are simulated to evaluate the potential and effectiveness of the proposed lightweight precision stage design method. 
Case study~\#1 considers a simple rib-enhanced stage structure with arbitrary sensor/actuator placements, aiming at demonstrating the impact of the selection of the weighting variable $\gamma$ on controller design. 
Case study~\#2 implements the proposed framework for a  practical lightweight planar motor  stage  with the actuator's weight and location constraints considered. The performance of both case studies  compared to that of  a baseline stage design without flexible mode control for evaluation. %More details in the numerical implementation will be discussed.

\subsection{Case study~\#1}

Figure~\ref{fig:case_1} shows the diagrams of the  stage structure being considered, which shows a rib-reinforced structure  made of 6061-T6 aluminum alloy of $300$~\rm{mm}$~\times~300$~\rm{mm} in size. The coordinate system being used is also shown in Fig.~\ref{fig:case_1}.  Herein, the rigid-body motion of the stage in three DOFs, including vertical translation ($z$), roll ($\theta_x$),  and pitch ($\theta_y$) are actively controlled. In addition, the proposed stage also actively controls its first vibration mode, and the baseline stage has no control for flexible modes. Therefore, three actuators and three sensors are used for the baseline stages for exact constraint, while the proposed case uses four actuators and four sensors. The geometric parameters  $\theta_p \in\mathbb{R}^{5}$  and the actuator/sensor location parameters $\theta_a = [x_a,y_a]^\top$ and $\theta_s=[x_s,y_s]^\top$ are also shown in Fig.~\ref{fig:case_1}. %In contrast, three sensors and three actuators are used for the baseline stage, which fully measures the rigid body motion of the stage in the considered DOFs.

Due to the geometric complexity of the ribbed stage structure, analytical models are not sufficient to capture its structural dynamics accurately. In this work, finite element (FE) simulation (with COMSOL Multiphysics) is used to simulate the stage's spatial-temporal behavior. 
In the stage geometry optimization problem \eqref{eqn: shape_opt} formulation for the proposed stage in Fig.~\ref{fig:case_1}, 
to facilitate controller design with a target control bandwidth of $\sim100~\rm{Hz}$, the values of $\omega_{low}$ and $\omega_{high}$ are selected as $50~\rm{Hz}$ and $500~\rm{Hz}$, respectively. In addition, the rib width and base height are constrained to be larger than $1~\rm{mm}$ for the sake of manufactuability. 
With the stage geometry optimization problem \eqref{eqn: shape_opt} fully formulated, the Optimization Module in COMSOL Multiphysics is selected to solve the problem, where  an iterative method for derivative-free constrained optimization COBYLA~\cite{powell1994direct} is employed. 
% The resultant optimal stage geometric parameters are $\theta = [1~mm,1~mm,65.87~mm,1~mm,12.74~mm]^\top$, with 
The resultant stage resonance frequencies and mode shapes are illustrated in Fig.~\ref{fig:case_1}.

% \begin{figure}[t!]
% \centering
% \subfloat{
% \includegraphics[trim={0mm 0mm 0mm 0mm},clip,width =0.8\columnwidth, keepaspectratio=true]{Figures/case_1_mode_all.png}}
% \vspace{-2mm}
% \caption{First four vibration mode shapes of case~\#1 }
% \vspace{-6mm}
% \label{fig:case_1_modes}
% \end{figure}

%%%% Actuator/sensor placement
The actuator/sensor placement optimization problems \eqref{eqn:act_opt}-\eqref{eqn:sen_opt} are then solved for the optimized structure. In case study \#1, the actuator/sensor location range is over the entire top surface of the stage, i.e., $D_a = D_s = \{(x,y,z) \big|~\|x\|,\|y\| \leq 0.15~\rm{m}, z=0\}$. 
%  $\Phi = [\phi_1 , \cdots , \phi_n]$ is an $n \times n$ matrix where $\phi_i$ represents the vector of corresponding $ith$ mode shape with mass matrix normalized
The normalized mode shapes over all mesh nodes $\phi_i(x,y,z)$ for the stage  and their corresponding natural frequencies $\omega_i$ can be obtained from the FE simulations. For each node location within placement domain, let $\theta_a$ or $\theta_s = (x,y,z)\in D_a$ or $D_s$ and thus the actuation/sensing matrices $B_a(\theta_a)$ or $C_s(\theta_s)$ can be found. A modal damping of $\zeta = 0.01$ is assumed for all modes, and the  grammians \eqref{eq:W_o} for each mode can be computed. A direct search algorithm is utilized to find the optimal actuator/sensor locations.  
%With these values, the controllability/observability grammiams \eqref{eq:W_p} and \eqref{eq:W_o} of the proposed stage can then be computed. 

\begin{rem}
When $\gamma$ and the placement domain of  actuators and sensors being identical, i.e., $D_a = D_s$, the optimal solution for both \eqref{eqn:act_opt} and \eqref{eqn:sen_opt} will be identical too. Therefore, the optimal configuration is a ``collocated'' case with the actuator and sensors configured at the same location \cite{rankers1998machine}. %This can be easily shown by taking the transpose of either numerator and using the property of the assembling matrices. 
In addition, the stage structure being considered is  symmetrical about the $x$ and $y$ axes. Therefore,  the optimal actuator/sensor location  will be also  symmetrical as shown in  Fig.~\ref{fig:case_1}. %This can also be proved by the fact that the magnitude of mode shapes of a square stage is symmetric about x and y axis (or by its dual mode with the same natural frequency). 
These two facts significantly simplify the numerical computation required for the actuator/sensor placement optimization problems.% ,without sacrificing any optimality in this case.
\end{rem}

With the stage's geometric design and the placement of actuator/sensor decided, we are able to extract the state-space models for the proposed lightweight stage from the FE simulations. 
%Fig.~\ref{fig:case_1_modes} shows the first four vibration modes of the stage obtained from Eigenfrequency study. 
The system's undamped dynamics can be written as 
\begin{align} \label{eqn:case_1_FE} 
\begin{split}
M_{FE}\ddot{x}_{FE}+K_{FE}x_{FE} &= B_{FE}u,
\\
    y &= C_{FE}x_{FE},
\end{split}
\end{align} 
where $x_{FE} \in\mathbb{R}^{n^{FE}}$ is the vector of displacement of all nodes in the FE simulation, $n_{FE}$ is the number of nodes from mesh setting, $M_{FE},K_{FE}\in\mathbb{R}^{n_{FE}\times n_{FE}}$ are the mass and stiffness matrices, respectively, and $B_{FE}$ and $C_{FE}$ are the input and output matrices determined by the actuator and sensor locations.
Note that the dimension of the FE-computed system dynamics \eqref{eqn:case_1_FE} is typically very large ($n_{FE}\sim 10^4$) especially when a fine mesh is used in the simulation. To overcome this problem, the system dynamics \eqref{eqn:case_1_FE} is transformed into the modal coordinate as 
\begin{align}
\begin{split}  \label{eqn:case_1_decoupled}
    \ddot{q} + Kq = B(\theta_a)u,  \\
    y = C(\theta_s)q,
\end{split}
\end{align}
where $q=\Phi^{-1}x_{FE}$ is the decoupled modal state vector, $\Phi = [\phi_1 , \cdots , \phi_n]$ is an $n \times n$ matrix where $\phi_i$ represents the vector of corresponding $i$-th mode shape with mass matrix normalized, i.e. $\Phi^\top M_{FE} \Phi = I$, $K = \Phi^\top K_{FE} \Phi$ is the diagonal stiffness matrix, and $B(\theta_a)$ and $C(\theta_s)$ are decoupled input and output matrix, respectively. 
In this decoupled coordinate, we can reduce the model order by truncating high-frequency vibration modes. We keep only the 3 rigid-body modes and first 10 flexible modes in this paper. Such model is able to capture the system dynamics accurately up to $1200~\rm{Hz}$, which is sufficient for controller design. Then, a modal damping term is introduced into the \eqref{eqn:case_1_decoupled}, and a reduced-order model in the form of \eqref{eqn:mech_EOM_1} can be derived. Finally, the actuation signals $u$ and measurement signals $y$ are transformed into the decouple DOFs as shown in Fig.~\ref{fig:control_diagram}.

\begin{figure}[t!]
\centering
\subfloat{
\includegraphics[trim={0mm 0mm 0mm 0mm},clip,width =0.5\columnwidth, keepaspectratio=true]{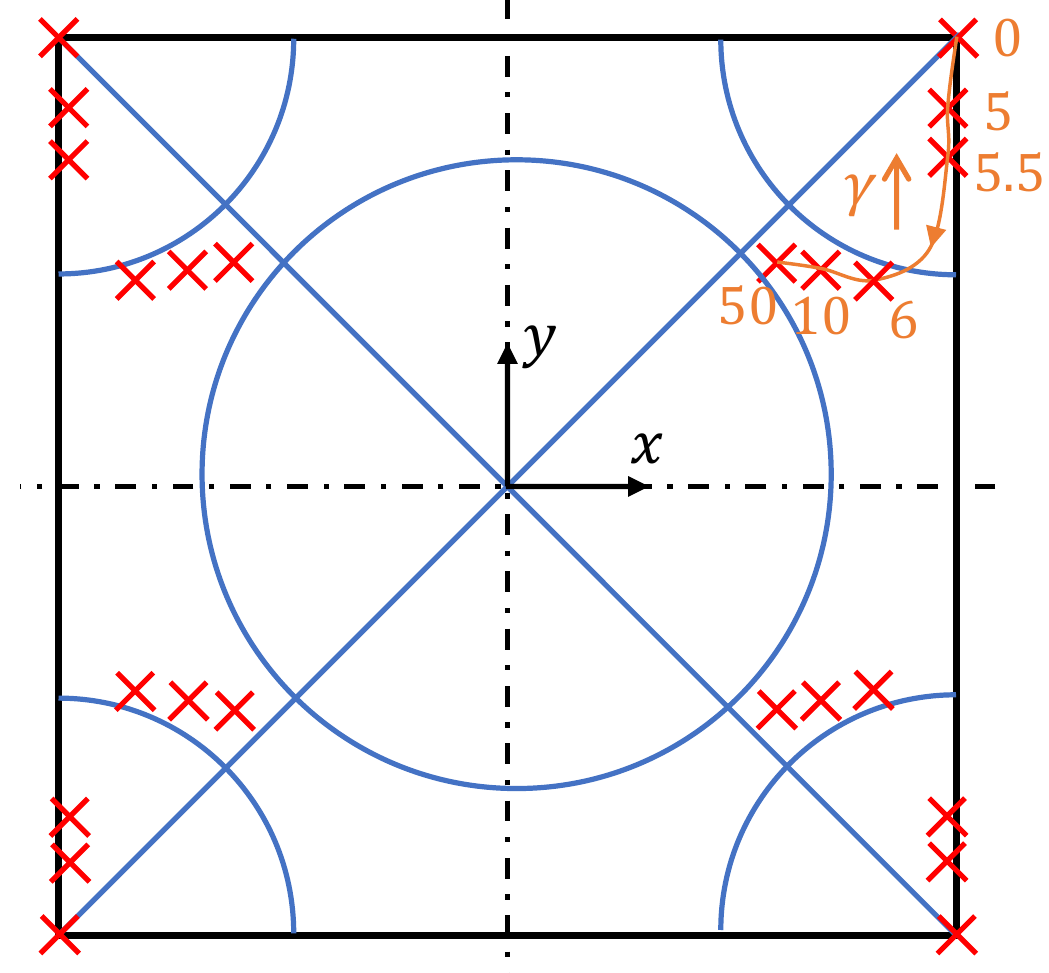}}
\vspace{-2mm}
\caption{Optimal actuator/sensor placements under varying $\gamma$. Blue: nodal points of uncontrolled modes. }
\vspace{-2mm}
\label{fig:case_1_gamma}
\end{figure}

\begin{figure}[t!]
\centering
\subfloat{
\includegraphics[trim={0mm 0mm 0mm 0mm},clip,width =1\columnwidth, keepaspectratio=true]{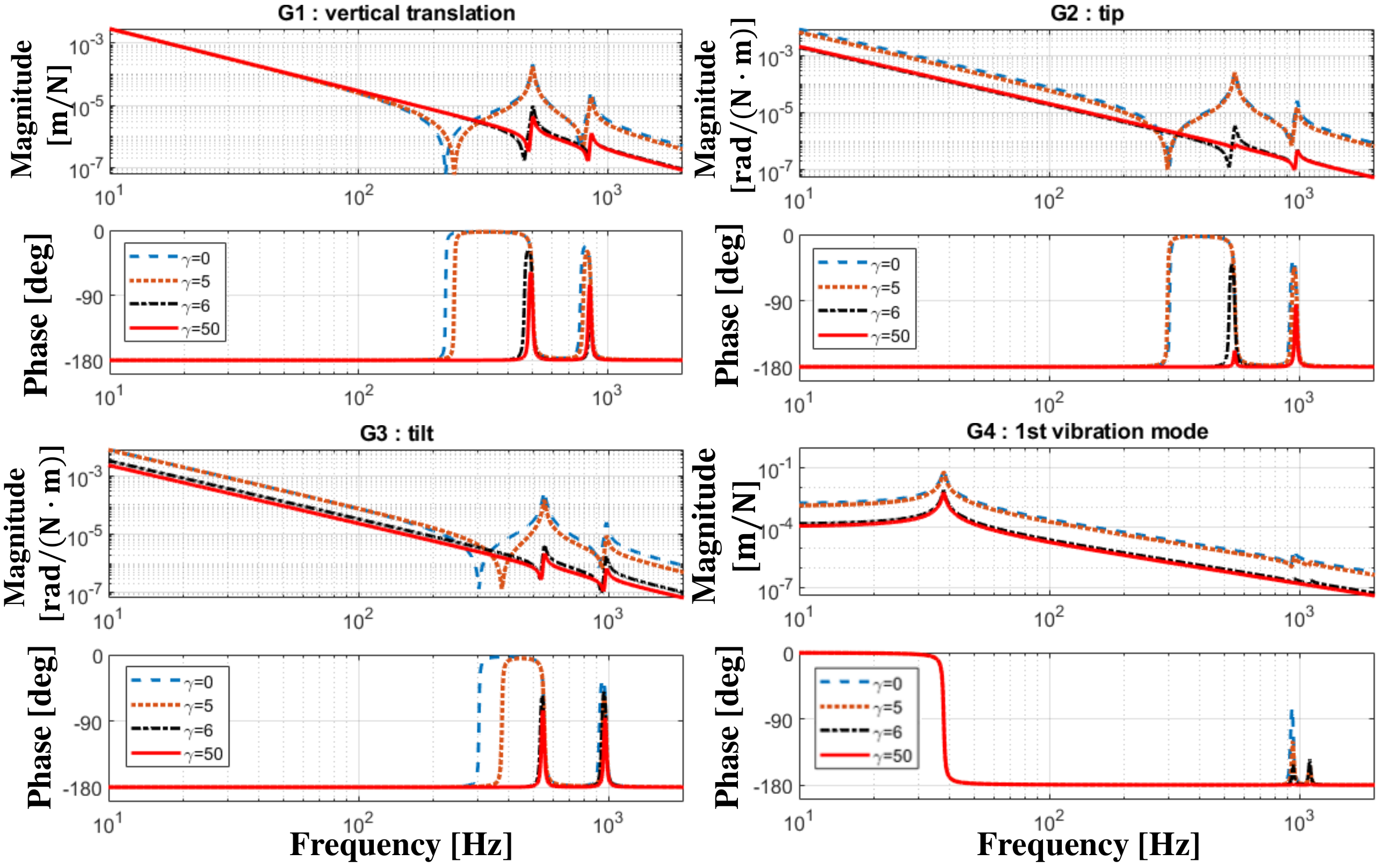}}
\vspace{-2mm}
\caption{Open-loop plant with different $\gamma$. }
\vspace{-4mm}
\label{fig:case_1_bode}
\end{figure}

As is stated in Section~\ref{sec:act_sen_placement},  the weighting parameter $\gamma$ in \eqref{eqn:act_opt}-\eqref{eqn:sen_opt} provides a balance between the need to have small control gains and the need to decouple controlled modes and uncontrolled modes. In this case study, \eqref{eqn:act_opt}-\eqref{eqn:sen_opt} are  solved for the stage structure with a varying value of $\gamma$, and the resultant actuator/sensor locations are shown in Fig.~\ref{fig:case_1_gamma}. Here in Fig.~\ref{fig:case_1_gamma}, the red crosses represent the optimal actuator/sensor locations with different $\gamma$ values (note that the actuators and sensors are collocated), and the blue lines represent the nodal lines of the stage's second to fourth vibration modes.
Figure~\ref{fig:case_1_bode} shows the decoupled plant frequency responses of the proposed lightweight stage with actuator/sensor location optimized under different values of $\gamma$. Several selections of the value $\gamma$ are discussed as below. 

\noindent \underline{(a): $\gamma = 0$}: With $\gamma = 0$, 
the optimal actuator/sensor locations are at the corners of the stage (Fig.~\ref{fig:case_1_gamma}), where the first vibration mode's modal displacement is maximized. This is because with $\gamma = 0$ we are only considering the need to maximize the controllability/observability of the actively-controlled modes, and not considering the effects of high-frequency uncontrolled modes. This is  confirmed by the plant frequency response shown in Fig.~\ref{fig:case_1_bode} with $\gamma =0$ (blue dashed line), where the last channel of the plant dynamics (the stage's first flexible mode) is having high magnitude. However, this design results in strong coupling between the stage's rigid body motion and the uncontrolled flexible modes (e.g. the second mode at 500~Hz). 

\begin{figure*}[t!]
\centering
\subfloat{
\includegraphics[trim={0mm 0mm 0mm 0mm},clip,width =1.5\columnwidth, keepaspectratio=true]{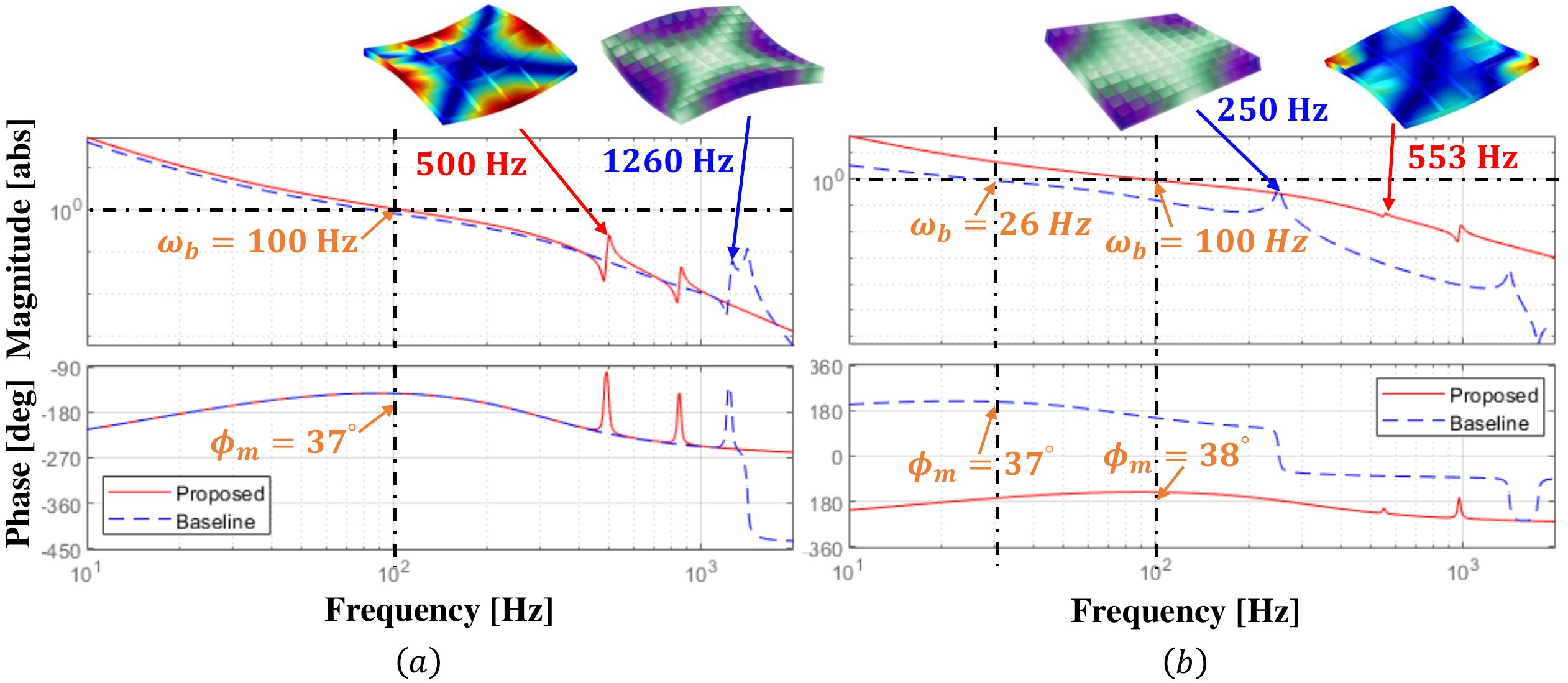}}
\vspace{-2mm}
\caption{Case study \#1: comparison of loop gains of the proposed stage design (red solid) and baseline stage design (blue dashed). (a) $z$-DOF (translation in the vertical direction).  (b)  $\theta_x$-DOF (pitch). }
\vspace{-4mm}
\label{fig:case_1_closed_loop}
\end{figure*}

\noindent \underline{(b): $\gamma = 50$}. As $\gamma$ increases, the actuator/sensor locations move towards the the nodal location of the stage's uncontrolled flexible modes, as shown in Fig.~\ref{fig:case_1_gamma}. This is also confirmed by the plant frequency responses shown in Fig.~\ref{fig:case_1_bode}: as $\gamma$ increases, the peak of uncontrolled flexible modes decreases, while the magnitude of the last channel in the plant dynamics (the stage's first flexible mode) reduces as well. 

From the discussions above, it can be concluded that a large value in $\gamma$ is beneficial for obtaining high control bandwidth at the cost of needing a higher controller gain in the flexible mode control. Therefore, the value of $\gamma$ should be selected as its maximum allowed value to produce an acceptable plant magnitude in the actively controlled flexible mode. In this case study, $\gamma = 50$ (i.e. the plant as red solid lines in Fig.~\ref{fig:case_1_bode}) is selected to enable a high control bandwidth. The resultant optimal actuator/sensor locations are close to the nodal positions of the uncontrolled flexible modes, see  Fig.~\ref{fig:case_1_gamma}. 
Finally, four SISO controllers in the form of \eqref{eqn:PID}  are designed for each actively-controlled DOFs, with a target control bandwidth of $\omega_{bw} = 100~\mathrm{Hz}$.

\begin{table}[t!] 
\centering
\renewcommand{\arraystretch}{1.0}
\caption{Case study \#1 performance comparison.}
\vspace{-4mm}
\label{table:case_1_para}
    \begin{center} \begin{small}
        \begin{tabular}{  p{2.7cm}  p{2.4cm}  p{2.4cm} }
        \hline
           & Baseline Design & Proposed Design \\ \hline
           Stage weight & 2.31 kg& 0.34 kg \\
           1st res. freq. & 250~Hz & 38 Hz  \\
           2nd res. freq.  & 1260~Hz & 500 Hz  \\
           $z$  bandwidth & 100 Hz  & 100 Hz \\
           $\theta_x/\theta_y$ bandwidth  & 26 Hz  & 100 Hz \\
        %   First res. freq. in CL & 250 Hz  & 100 Hz  \\
        %   First res. damp. ratio & 0.01 & 0.32 \\
           Max sensitivity  & 1.89  & 1.84 \\ 
        \hline
        \end{tabular}
            \vspace{-8mm}

    \end{small} \end{center}
\end{table}

To evaluate the effectiveness of our proposed design method, a baseline lightweight precision stage as illustrated in Fig.~\ref{fig:case_1}  is used for comparison. This baseline stage lightweight stage does not have active control for its flexible modes, and only has the rigid body motions under feedback control. Three actuators and three sensors are used to achieve exact constraint in the stage actuation and control. In such a design, the first resonance frequency of the  stage structure places an upper limit to the achievable control bandwidth. With a target control bandwidth of 50~Hz, the geometric parameters of the baseline stage are designed such that the first resonance frequency of the stage structure is above 250~Hz (i.e. $5\times$ of the target bandwidth). Similarly, SISO controllers in the form of \eqref{eqn:PID} are designed for all decoupled DOFs under active control such that the robustness criteria  \ref{eqn:robustness} is satisfied.

% \lei{Please read the following three paragraphs as an example of how to discuss figures in the evaluation section of a paper. Keep in mind that NOTHING IS OBVIOUS to the reader and you must deliver what you want to show to the reader in a CLEAR, DETAILED, and DIRECT manner.}

Table.~\ref{table:case_1_para} summarizes the performance of the proposed lightweight stage in case study~\#1 and that of the baseline stage,  and Fig.~\ref{fig:case_1_closed_loop} shows the loop gains of both proposed and baseline designs in the $z$-DOF (translation in the vertical direction) and the $\theta_x$-DOF (pitch direction).  
Comparing the loop frequency responses shown in Fig.~\ref{fig:case_1_closed_loop}a, it can be observed both stages can reach a high control bandwidth of $100~\rm{Hz}$ with sufficient stability margins in the $z$-DOF, and the $250$~Hz resonance in the baseline stage is not shown in its $z$-DOF frequency response. This is because the baseline's first flexible mode is not controllable or not excitable by the $z$-axis control loop, and thus this resonance does not limit the stage's control bandwidth in this axis. However, the $250$~Hz resonance of the baseline stage can couple in the stage's $z$-axis dynamics under imperfect actuator or position placement, and stability issue can arise in the control under such situations. In addition, the lightly-damped resonance at $250$~Hz in the baseline stage is not actively controlled and thus can be easily excited by disturbances, which can impair the stage's positioning accuracy.

Comparing the loop frequency responses shown in Fig.~\ref{fig:case_1_closed_loop}b, it can be observed that the bandwidth of the baseline stage is only  26~Hz. This is primarily due to the 250~Hz resonance peak in the stage dynamics is coupled into the stage's control in the $\theta_x$ direction with the current actuator/sensor configuration, and thus limits the achievable control bandwidth. In contrast, the proposed design can robustly achieve a control bandwidth of 100~Hz since the stage's first resonance mode at 50Hz is actively controlled. 

Finally, comparing the performance shown in Table~\ref{table:case_1_para}, it can be seen  that the  weight of the proposed stage design is reduced by 85\% compared to baseline design. To our understanding, this significant gain in weight reduction is due to the proposed stage is allowing compliance in the first flexible mode, which effectively removes material in the stage structure needed to reinforce the stage. This result shows the tremendous potential of the proposed approach in stage acceleration improvement and the power consumption reduction. 
In addition, comparing the closed-loop damping performance of the stage's first resonance mode, it can be seen that the baseline stage's resonance at $250$~Hz is only having a low damping ratio of 0.01, which can be excited by external disturbances. In contrast, the first flexible mode of the proposed stage is under closed-loop control, which has a bandwidth of $100$~Hz and has a closed-loop damping ratio of $0.37$. This improvement in the structural damping shows the potential of the proposed approach to improve the stage's positioning accuracy under external disturbances.

% However, in tip motion channel (b), our proposed design can still achieve the target bandwidth while the baseline design only has 26~Hz control bandwidth mainly limited by the 250~Hz resonance peak which is the first flexible mode.

% Note that from the loop gain of proposed design (a) in Fig.~\ref{fig:case_1_closed_loop}, we can clearly see that there is still a big gap between crossover bandwidth frequency and the 500~Hz $2$nd resonance mode and also between its peak and unit magnitude, which motivates the reduction on $\omega_{high}$ to remove the conservatism.

% \begin{table}[t!] 
% \centering
% \renewcommand{\arraystretch}{1.0}
% \caption{Optimal parameters}
% \vspace{-4mm}
% \label{table:case_1_para}
%     \begin{center} \begin{small}
%         \begin{tabular}{ | m{3.5em} | m{1cm} | m{0.9cm}| m{1cm}| m{1.3cm} | m{1.2cm} |}
%         \hline
%           & Weight & 1st res. freq. & 2nd res. freq. & Bandwidth  & Sensitivity peak \\
%         \hline
%         Proposed Design & 0.34~kg &  38~Hz & 500~Hz & 100~Hz  &  1.84 \\
%         \hline
%         Baseline  & 2.31~kg & 250~Hz & 1260~Hz & 26~Hz  & 1.89 \\
%         \hline

%         \end{tabular}
%     \end{small} \end{center}
% \end{table}

\subsection{Case study~\#2}\label{sec:case_study_2}
\begin{figure*}[t!]
\centering
\subfloat{
\includegraphics[trim={0mm 0mm 0mm 0mm},clip,width =2\columnwidth, keepaspectratio=true]{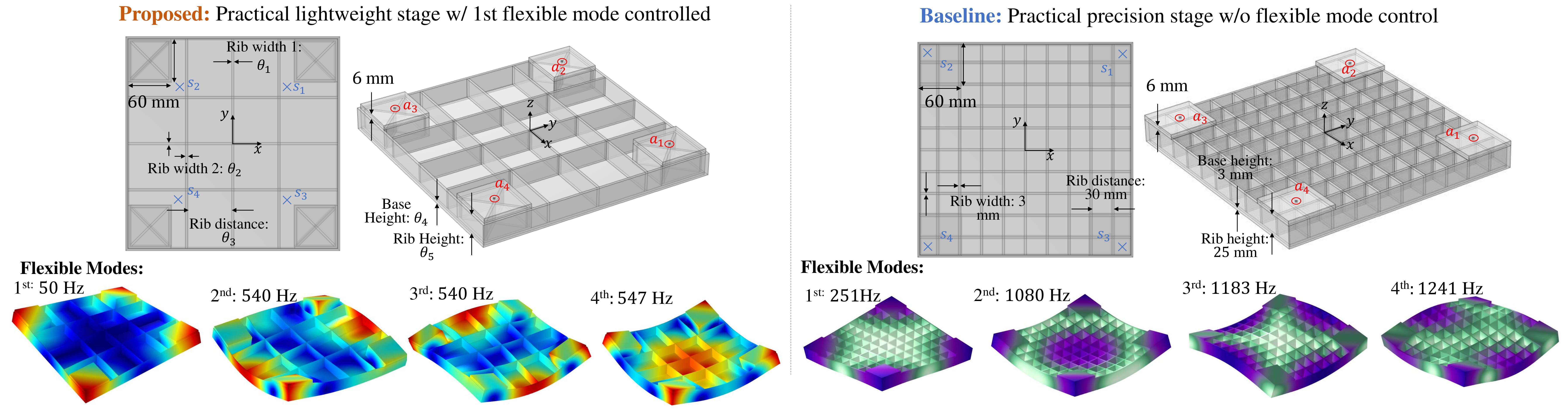}}
\vspace{-2mm}
\caption{Case study \#2 proposed and baseline stages. Both stages consider a permanent magnet array with $60~\mathrm{mm}\times60~\mathrm{mm}\times6~\mathrm{mm}$ for planar motor force generation. }
\vspace{-4mm}
\label{fig:case_2_definition}
% \end{figure*}
% \begin{figure*}[t!]
\centering
\subfloat{
\includegraphics[trim={0mm 0mm 0mm 0mm},clip,width =1.5\columnwidth, keepaspectratio=true]{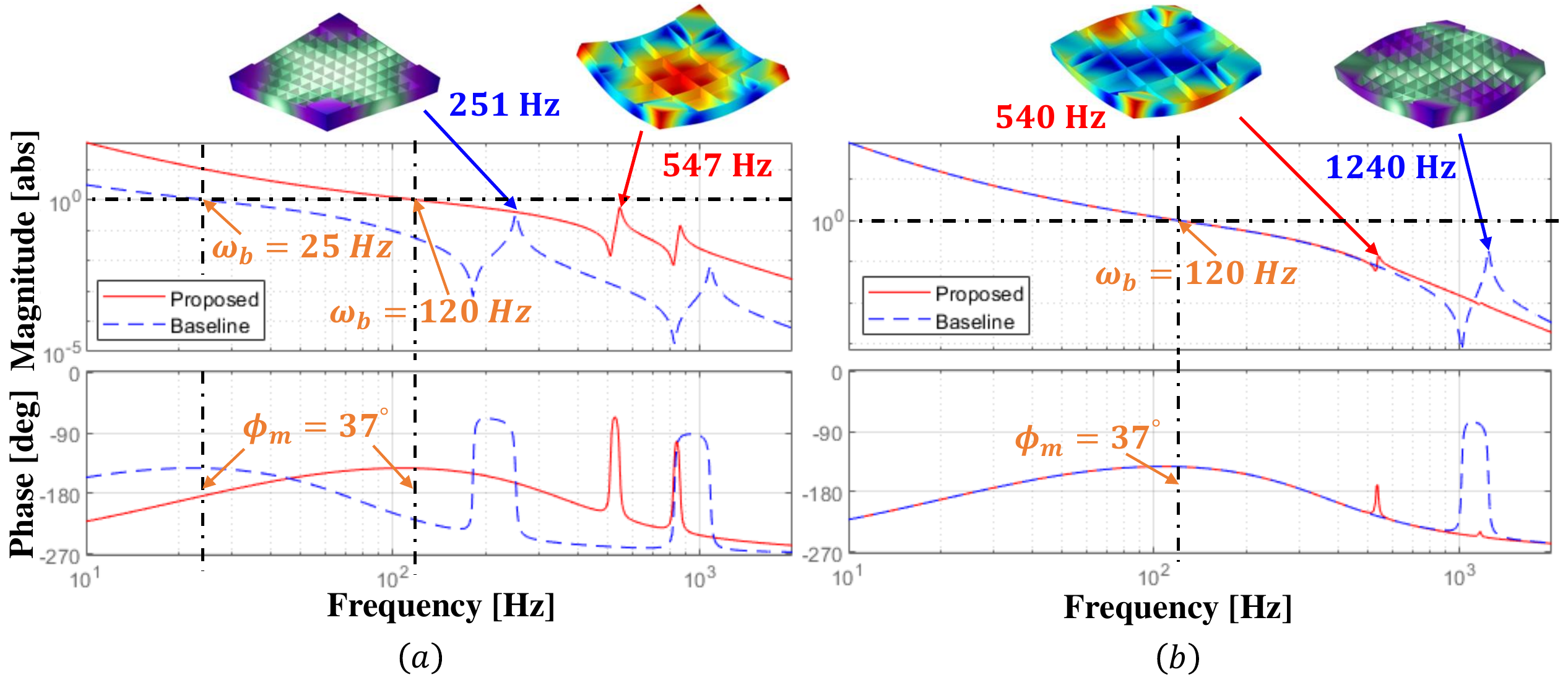}}
\vspace{-2mm}
\caption{Case study \#2: comparison of loop gains of the proposed stage design (red solid) and baseline stage design (blue dashed). (a) $z$-DOF (translation in the vertical direction).  (b)  $\theta_x$-DOF (pitch). }
\vspace{-3mm}
\label{fig:case_2_closed_loop}
\end{figure*}

Case study~\#2 considers a magnetically-levitated planar motion stage as illustrated in Fig.~\ref{fig:case_2_definition}, where four neodymium permanent magnet arrays of $60\mathrm{mm} \times 60~\mathrm{mm}\times 6~\mathrm{mm}$ are arranged at the corner of the stage to provide both thrust forces for planar motion and the levitation forces. The inclusion of the actuator magnets  enhances the practical relevance of the case study for wafer positioning application. The vertical-directional levitation forces are assumed to be located at the center of the permanent magnet arrays. All other stage geometry parameters are defined in the same way with case study~\#1.

As stated in Remark~\ref{rem:1}, the value of $\omega_{high}$ sets an upper bound for the achievable control bandwidth for the proposed positioning stage. %with actively controlled flexible modes. 
However, using a high value of   $\omega_{high}$  can enforce the stage design to increase materials to stiffen the corresponding resonance mode, and thus increase the stage's weight. Therefore, to fully explore the feasible designs set as illustrated in Fig.~\ref{fig:motivation} and thus to remove possible design conservatism, the value of  $\omega_{high}$  needs to be swept. It is worth pointing out that the stage geometry design \eqref{eqn: shape_opt} and the actuator/sensor placement design \eqref{eqn:act_opt}-\eqref{eqn:sen_opt} collaboratively determine the plant dynamics of the positioning stage. When conducting a parameter sweep for $\omega_{high}$, the actuator/sensor placement problems must also be solved for each stage geometry design for effective design optimization. 

% \begin{rem}
% The stage's weight obtained from the solution of \eqref{eqn: shape_opt} must be lower than or at least equal to that from  \eqref{eqn: shape_opt} with the same $\omega_{low}$ and a larger $\omega_{high}$ since the optimal solution of previous one is always feasible in the latter program.
% \end{rem}

To reduce possible design conservatism and thus fully exploit the advantages brought by the flexible mode control, the feasible stage design set for case study \#2 is explored as follows:  First, a target control bandwidth is selected to be $120$~Hz for the positioning stage. Next,  the stage geometry optimization problem \eqref{eqn: shape_opt} is solved with   $\omega_{high} = 600~\mathrm{Hz}$, i.e. $5\times$ of the target bandwidth. Then, the sensor positioning optimization problem \eqref{eqn:sen_opt} is solved with $\gamma = 50$. Note that the actuator's locations are fixed due to the inclusion of magnet arrays.  With one feasible stage and sensor positioning design provided by the previous steps, we then decrease the value of $\omega_{high}$ by a constant step $\delta \omega = 10~\mathrm{Hz}$ and resolve \eqref{eqn: shape_opt} and \eqref{eqn:sen_opt}. Assuming $\delta \omega$ is sufficiently small, the change in optimal geometric  parameters can be assumed continuous, which allows us to  use the  optimal solution from the previous run as the initial parameters when resolving \eqref{eqn: shape_opt}. This method effectively reduces the required computation time. %After obtaining the stage's geometry in each step, the sensor placement optimization \eqref{eqn:sen_opt} is also resolved with $\gamma=50$. 
The previous steps are repeated until  $\omega_{high}$ is sufficiently low such that it may be excited by external disturbances. In this case study, the lowest value of  $\omega_{high}$  is selected to be at 300~Hz.

\begin{figure}[t!]
\centering

\includegraphics[trim={0mm 0mm 0mm 0mm},clip,width =1\columnwidth, keepaspectratio=true]{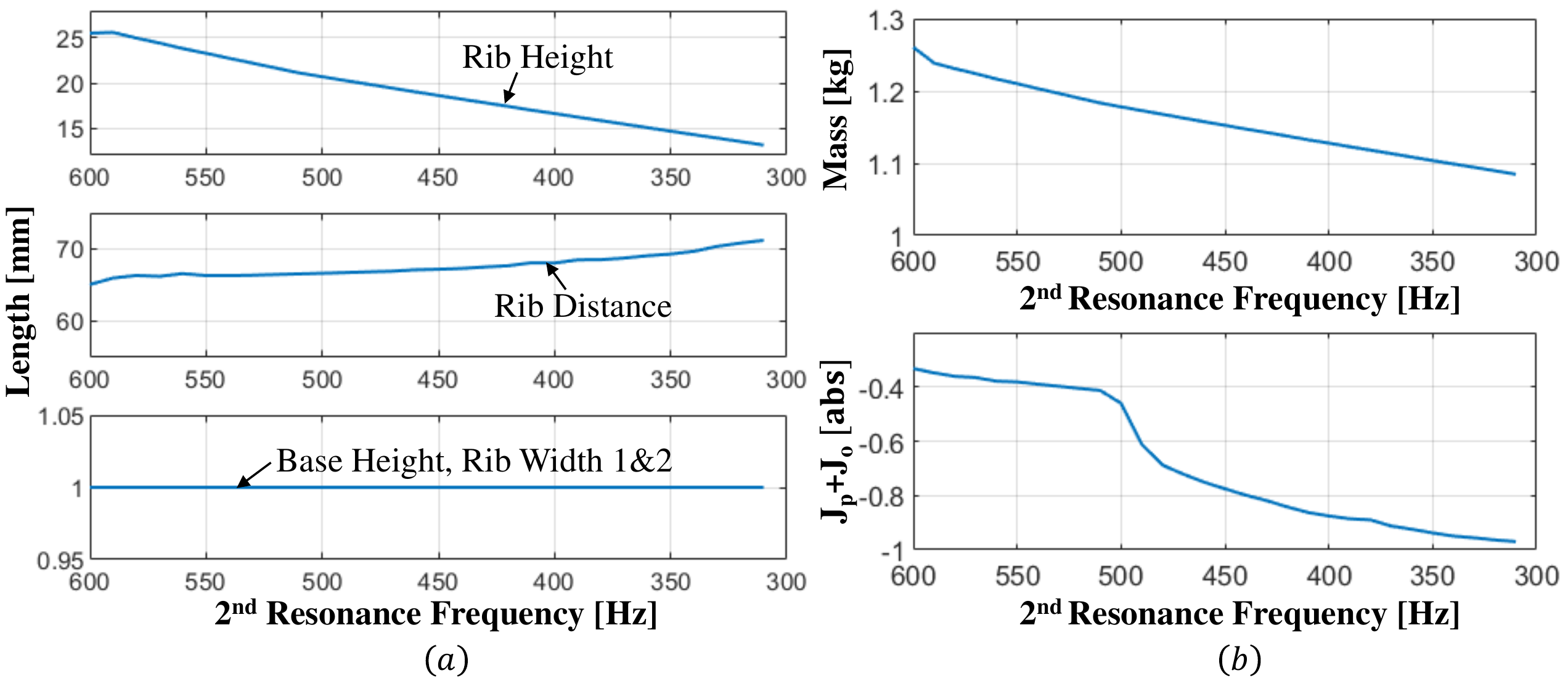}
% \vspace{-2mm}
% \caption{ Geometric parameter history. }
% \vspace{-2mm}
% \label{fig:case_2_para_hist}
% \end{figure}
% \begin{figure}[t!]
% \includegraphics[trim={0mm 0mm 0mm 0mm},clip,width =1\columnwidth, keepaspectratio=true]{Figures/case_2_both.png}
\vspace{-8mm}
\caption{ (a) Geometric parameter history. (b) Stage weight and grammian history. }
\vspace{-6mm}
\label{fig:case_2_mass_gram}
\end{figure}

% \begin{figure}[t!]
% \centering
% \subfloat{
% \includegraphics[trim={0mm 0mm 0mm 0mm},clip,width =0.8\columnwidth, keepaspectratio=true]{Figures/normalized_pareto_front.png}}
% \vspace{-2mm}
% \caption{ Pareto front. }
% \vspace{-2mm}
% \label{fig:case_2_pareto}
% \end{figure}

In the stage geometry optimization problem, the optimal solutions always have the stage's second resonance frequency match $\omega_{high}$. 
Fig.~\ref{fig:case_2_mass_gram} shows the stage geometric parameters and the  resultant stage weight and actuator/sensor placement objectives under varying $\omega_{high}$. 
%Meanwhile, all the intermediate parameters, natural frequencies, optimal objective values from \eqref{eqn:act_opt} and \eqref{eqn:sen_opt} and the stage's weights are recorded in Fig.~\ref{fig:case_2_para_hist} and Fig.~\ref{fig:case_2_mass_gram}.
It can be observed that %as  $\omega_{high}$ decreases, the stage's rib height decreases,  while the variation of other parameters are small. In addition, 
the stage's weight is reducing as the value of $\omega_{high}$ decreases, and the value of $J_p + J_o$ (i.e.~sum of objectives of \eqref{eqn:act_opt}-\eqref{eqn:sen_opt}) is also decreasing along with the reduction of $\omega_{high}$. These observations reveal new trade-off between the stage's achievable control bandwidth and acceleration (assuming constant thrust force generation), which is illustrated by the orange line in Fig~\ref{fig:motivation}.

The stage hardware design can be manually made among the optimal designs based on the results shown in Fig.~\ref{fig:case_2_mass_gram}. In this case study,  $\omega_{high} = 540~\mathrm{Hz}$ is selected to provide sufficiently high $J_p+J_o$ values while reducing the stage's weight. Compared to the initial stage design using $\omega_{high} = 600~\mathrm{Hz}$, the stage's weight is reduced by 4.5\%. Although the improvement is not significant, it is worth pointing out that the geometry optimization of the stage is relatively limited in the current formulation with only five parameters that can be varied. A more significant improvement in the stage's performance may be expected given increased design flexibility is allowed in the stage structure. 
% The improvement is not too much since there is certain limitation of only controlling first flexible mode and leaving the rest uncontrolled in current geometric parametrization. If we want to further reduce the weight to a large extent, we need to go back to the general framework \eqref{eqn: shape_opt} and enlarge $n$ to exploit more potential of the stage topology. \jingjie{The preceding 2 sentences may be moved to future work}.
The resultant stage's flexible modes are illustrated in the bottom left in Fig.~\ref{fig:case_2_definition}. The state-space dynamic  model of the stage can be derived for this stage in the same way as discussed in case study~\#1, and controllers are designed for the decoupled motions. 

% To evaluate the effectiveness of proposed framework and the parameter sweeping process, a baseline conventional realistic stage design as shown in Fig.~\ref{fig:case_2_para} (c) and (d) is simulated for comparison. Similarly, four actuators and sensors are fixed at given locations and we decoupled the system into only 3 rigid-body motions. Table.~\ref{fig:case_2_mass_gram} summarizes the performance and comparisons. The proposed stage design's weight is reduced by 55\% compared to the baseline design. The control design in proposed case can reach target bandwidth in all channels while the baseline design can only achieve 25~Hz control bandwidth in vertical translation motion due to the 250~Hz resonance peak limitation. In both cases, the robustness criteria \eqref{eqn:robustness} is strictly satisfied.

% Figure.~\ref{fig:case_2_closed_loop} shows the loop gains of both proposed and baseline designs. Clearly, we can see that in vertical translation channel (a),  both designs reach a good phase margin and the first resonance mode in baseline design is greatly limiting the bandwidth while in proposed case, the $3$rd mode with 547~Hz is now the limiting factor. In tip channel (b), both designs can achieve the target bandwidth and phase margin since in baseline case, the first resonance mode is not showing up due to decoupling.

%
To evaluate the effectiveness of our proposed framework considering actuator weight and constraints, a baseline lightweight stage with same magnet array is simulated for comparison. In the baseline stage, only the rigid-body motions are under active control, and all flexible modes are  uncontrolled.  %Accordingly, four actuators with actuating forces located at the center of PM arrays and four fixed sensors are used to decouple the system into the 3 rigid DOFs and provide closed-loop feedback control for each. 
With a target bandwidth of 50~Hz, the stage's geometric parameters are designed to constrain the first resonance frequency above 250~Hz. Fig.~\ref{fig:case_2_definition} show the baseline stage design parameters and actuator/sensor location. Three SISO controllers as \eqref{eqn:PID} are designed for all decoupled DOFs in the same way with case study~\#1.

\begin{table}[t!] 
\centering
\renewcommand{\arraystretch}{1.0}
\caption{Case~\#2 Optimal parameters}
\vspace{-4mm}
\label{table:case_2_para}
    \begin{center} \begin{small}
        \begin{tabular}{  p{2.7cm} | p{2.4cm} | p{2.4cm} }
        \hline
           & Baseline Design & Proposed Design \\ \hline
           Stage weight & 2.67 kg& 1.20 kg \\
           First res. freq. & 251~Hz & 50 Hz  \\
           2nd res. freq.  & 1080~Hz & 540 Hz  \\
           $z$ motion bandwidth & 25 Hz  & 120 Hz \\
           $\theta_x/\theta_y$ bandwidth  & 120 Hz  & 120 Hz \\
        %   First res. freq. in CL & 251 Hz  & 120 Hz  \\
        %   First res. damp. ratio & 0.01 & 0.32 \\
           Max sensitivity  & 1.80  & 1.94 \\ 
        \hline
        \end{tabular}
            \vspace{-8mm}
    \end{small} \end{center}
\end{table}

Table.~\ref{table:case_2_para} summarizes the performance and comparison of the proposed and baseline stage design in case~\#2,  Fig.~\ref{fig:case_2_closed_loop} illustrates the loop gains of both proposed and baseline designs in $z$- and  $\theta_x$-DOFs. Comparing the loop frequency responses in Fig.~\ref{fig:case_2_closed_loop}a, it can be observed that the bandwidth of the baseline design is limited to 25~Hz due to the $251$~Hz resonance peak. In contrast, the proposed design can reach a bandwidth of 120~Hz with sufficient stability margin. % since the first resonance peak at 50Hz is actively controlled. 
Fig.~\ref{fig:case_2_closed_loop}b shows that both designs can reach a  bandwidth of 120~Hz  in the $\theta_x$-DOF. This is because the $251$~Hz resonance peak in the baseline stage is not excitable by the $\theta_x$ feedback loop. However, similar to the $z$-DOF in case study~\#1, stability issue can be caused if the actuator/sensor placement is imperfect. Moreover, the lightly-damped $251$~Hz resonance mode can be easily excited by external disturbance and thus  impair the stage's positioning precision.

Finally, Table~\ref{table:case_2_para} shows that the weight of the proposed stage design is reduced by 55\% compared to baseline design. The significant improvement for a stage considering the weight of magnet array shows the effectiveness and generality of our proposed approach. 
In addition,  comparing the closed-loop damping performance of stage's first resonance mode, it can be stated that the proposed design is more robust against external disturbances with the first lightly-damped mode at 547~Hz, while that of the baseline stage is at $251$~Hz. The comparison indicates the huge potential of our framework to improve both the stage's acceleration capability and positioning accuracy simultaneously.

% \begin{figure}[t!]
% \centering
% \subfloat{
% \includegraphics[trim={0mm 0mm 0mm 0mm},clip,width =1\columnwidth, keepaspectratio=true]{Figures/case_2_parameters.png}}
% \vspace{-2mm}
% \caption{First four vibration mode shapes of case~\#2. }
% \vspace{-2mm}
% \label{fig:case_2_parameters}
% \end{figure}

% \begin{figure}[t!]
% \centering
% \subfloat{
% \includegraphics[trim={0mm 0mm 0mm 0mm},clip,width =0.8\columnwidth, keepaspectratio=true]{Figures/case_2_mode_all.png}}
% \vspace{-2mm}
% \caption{First four vibration mode shapes of case~\#2. }
% \vspace{-2mm}
% \label{fig:case_2_modes}
% \end{figure}

%%%%% Conclusions and future work %%%%%%%%%%%%%%%%%%%%%%%%%%%%%%%

\section{Conclusion and future work}  \label{sec:conclusion}

In this work, we proposed and evaluated a sequential hardware and controller co-design framework for lightweight precision stages, aiming at enabling designs that can achieve high control bandwidth and high acceleration simultaneously. The algorithm of the framework is presented, and the effectiveness of the proposed method is demonstrated by numerical simulations using  two case studies. The significant weight reduction ($>$55\%) and improvement in control bandwidth show the potential. Future work will consider the  experimental evaluations for the proposed method. A fully integrated controller and hardware co-optimization that can better exploit the synergy between hardware and control designs will also be studied.

\balance

\bibliographystyle{IEEEtranS}
\bibliography{references}  %%% Uncomment this line and comment out the ``thebibliography'' section below to use the external .bib file (using bibtex).

% Generated by IEEEtranS.bst, version: 1.14 (2015/08/26)
\begin{thebibliography}{10}
\providecommand{\url}[1]{#1}
\csname url@samestyle\endcsname
\providecommand{\newblock}{\relax}
\providecommand{\bibinfo}[2]{#2}
\providecommand{\BIBentrySTDinterwordspacing}{\spaceskip=0pt\relax}
\providecommand{\BIBentryALTinterwordstretchfactor}{4}
\providecommand{\BIBentryALTinterwordspacing}{\spaceskip=\fontdimen2\font plus
\BIBentryALTinterwordstretchfactor\fontdimen3\font minus
  \fontdimen4\font\relax}
\providecommand{\BIBforeignlanguage}[2]{{%
\expandafter\ifx\csname l@#1\endcsname\relax
\typeout{** WARNING: IEEEtranS.bst: No hyphenation pattern has been}%
\typeout{** loaded for the language `#1'. Using the pattern for}%
\typeout{** the default language instead.}%
\else
\language=\csname l@#1\endcsname
\fi
#2}}
\providecommand{\BIBdecl}{\relax}
\BIBdecl

\bibitem{albero2011micromachined}
J.~Albero, S.~Bargiel, N.~Passilly, P.~Dannberg, M.~Stumpf, U.~Zeitner,
  C.~Rousselot, K.~Gastinger, and C.~Gorecki, ``Micromachined array-type mirau
  interferometer for parallel inspection of mems,'' \emph{Journal of
  Micromechanics and Microengineering}, vol.~21, no.~6, p. 065005, 2011.

\bibitem{butler2011position}
H.~Butler, ``Position control in lithographic equipment [applications of
  control],'' \emph{IEEE Control Sys. Mag.}, vol.~31, no.~5, pp. 28--47, 2011.

\bibitem{delissen2020high}
A.~Delissen, D.~Laro, H.~Kleijnen, F.~van Keulen, and M.~Langelaar,
  ``High-precision motion system design by topology optimization considering
  additive manufacturing,'' in \emph{20th Int. Conf. of the European Society
  for Precision Eng. and Nanotech., EUSPEN 2020}.\hskip 1em plus 0.5em minus
  0.4em\relax EUSPEN, 2020, pp. 257--258.

\bibitem{ding2020optimal}
R.~Ding, C.~Ding, Y.~Xu, W.~Liu, and X.~Yang, ``An optimal actuator placement
  method for direct-drive stages to maximize control bandwidth,'' in
  \emph{IECON 2020 The 46th Annual Conference of the IEEE Industrial
  Electronics Society}.\hskip 1em plus 0.5em minus 0.4em\relax IEEE, 2020, pp.
  556--561.

\bibitem{franklin2002feedback}
G.~F. Franklin, J.~D. Powell, A.~Emami-Naeini, and J.~D. Powell, \emph{Feedback
  control of dynamic systems}.\hskip 1em plus 0.5em minus 0.4em\relax Prentice
  hall Upper Saddle River, 2002, vol.~4.

\bibitem{garcia2019control}
M.~Garcia-Sanz, ``Control co-design: an engineering game changer,''
  \emph{Advanced Control for Appl.: Eng. and Ind. Sys.}, vol.~1, no.~1, p. e18,
  2019.

\bibitem{laro2010design}
D.~A. Laro, R.~Boshuisen, and J.~van Eijk, ``Design and control of
  over-actuated lightweight 450 mm wafer chuck,'' in \emph{2010 ASPE Spring
  Topical meeting, Cambridge, Massachusetts, USA}.\hskip 1em plus 0.5em minus
  0.4em\relax ASPE, 2010, pp. 141--144.

\bibitem{oomen2018advanced}
T.~Oomen, ``Advanced motion control for precision mechatronics: Control,
  identification, and learning of complex systems,'' \emph{IEEJ Journal of Ind.
  Appl.}, vol.~7, no.~2, pp. 127--140, 2018.

\bibitem{oomen2013connecting}
T.~Oomen, R.~van Herpen, S.~Quist, M.~van~de Wal, O.~Bosgra, and M.~Steinbuch,
  ``Connecting system identification and robust control for next-generation
  motion control of a wafer stage,'' \emph{IEEE Trans. on Ctrl. Sys. Tech.},
  vol.~22, no.~1, pp. 102--118, 2013.

\bibitem{ortega2004systematic}
M.~Ortega and F.~Rubio, ``Systematic design of weighting matrices for the h
  mixed sensitivity problem,'' \emph{Journal of Process Control}, vol.~14,
  no.~1, pp. 89--98, 2004.

\bibitem{powell1994direct}
M.~J. Powell, ``A direct search optimization method that models the objective
  and constraint functions by linear interpolation,'' in \emph{Adv. in opt. and
  num. analysis}.\hskip 1em plus 0.5em minus 0.4em\relax Springer, 1994, pp.
  51--67.

\bibitem{rankers1998machine}
A.~M. Rankers, ``Machine dynamics in mechatronic systems: An engineering
  approach.'' 1998.

\bibitem{van2017integrating}
G.~van~der Veen, M.~Langelaar, S.~van~der Meulen, D.~Laro, W.~Aangenent, and
  F.~van Keulen, ``Integrating topology optimization in precision motion system
  design for optimal closed-loop control performance,'' \emph{Mechatronics},
  vol.~47, pp. 1--13, 2017.

\bibitem{van2014exploiting}
R.~van Herpen, T.~Oomen, E.~Kikken, M.~van~de Wal, W.~Aangenent, and
  M.~Steinbuch, ``Exploiting additional actuators and sensors for
  nano-positioning robust motion control,'' \emph{Mechatronics}, vol.~24,
  no.~6, pp. 619--631, 2014.

\bibitem{JingjieCoDesign}
J.~Wu and L.~Zhou, ``Control co-design of actively controlled lightweight
  structures for high-acceleration precision motion systems,'' in \emph{2022
  American Control Conference (ACC)}, 2022, pp. 5320--5327.

\end{thebibliography}

\end{document}